\documentclass[twocolumn,superscriptaddress,prc,showpacs,amsmath,amssymb,letterpaper]{revtex4-1}
\usepackage{graphicx,color}
\usepackage{amssymb}
\usepackage{amsmath}
\usepackage{latexsym}
\usepackage{amsxtra}
\usepackage{amscd}
\usepackage{dcolumn}
\usepackage{bm}
\usepackage[breaklinks]{hyperref}
\usepackage{url}
\usepackage{breakurl}

\graphicspath{{./}}
\let\oldsqrt\sqrt
\def\sqrt{\mathpalette\DHLhksqrt}
\def\DHLhksqrt#1#2{\setbox0=\hbox{$#1\oldsqrt{#2\,}$}\dimen0=\ht0
\advance\dimen0-0.2\ht0
\setbox2=\hbox{\vrule height\ht0 depth -\dimen0}%
{\box0\lower0.4pt\box2}}

\newcommand{\nuc}[2]{$^{#1}$#2}

\begin{document}
\title{Isomeric states in neutron-rich nuclei around \boldmath{$N=40$}}

\author{K.~Wimmer}
\email{Corresponding author: k.wimmer@csic.es}
\affiliation{Instituto de Estructura de la Materia, CSIC, E-28006 Madrid, Spain}
\affiliation{Department of Physics, The University of Tokyo, 7-3-1 Hongo, Bunkyo-ku, Tokyo 113-0033, Japan}
\affiliation{RIKEN Nishina Center, 2-1 Hirosawa, Wako, Saitama 351-0198, Japan}
\author{F.~Recchia}
\author{S.~M.~Lenzi}
\affiliation{Dipartimento di Fisica e Astronomia dell' Universit\`a di Padova, Padova I-35131, Italy}  
\affiliation{INFN, Sezione di Padova, Padova I-35131, Italy} 
\author{S.~Riccetto}
\affiliation{Dipartimento di Fisica e Geologia dell' Universit\`a di Perugia, Perugia, Italy}
\affiliation{INFN, Sezione di Perugia, Perugia, Italy}
\author{T.~Davinson}
\affiliation{School of Physics and Astronomy, The University of Edinburgh, James Clerk Maxwell Building,  Edinburgh EH9 3FD, United Kingdom}
\author{A.~Estrade}
\affiliation{Central Michigan University, USA}
\author{C.~J.~Griffin}
\affiliation{School of Physics and Astronomy, The University of Edinburgh, James Clerk Maxwell Building,  Edinburgh EH9 3FD, United Kingdom}
\author{S.~Nishimura}
\affiliation{RIKEN Nishina Center, 2-1 Hirosawa, Wako, Saitama 351-0198, Japan}
\author{V.~Phong}
\affiliation{RIKEN Nishina Center, 2-1 Hirosawa, Wako, Saitama 351-0198, Japan}
\affiliation{Faculty of Physics, VNU University of Science, 334 Nguyen Trai, Thanh Xuan, Hanoi, Vietnam}
\author{P.-A.~S\"oderstr\"om}
\affiliation{RIKEN Nishina Center, 2-1 Hirosawa, Wako, Saitama 351-0198, Japan}
\author{O.~Aktas}
\affiliation{KTH Royal Institute of Technology, Stockholm, Sweden}
\author{M.~Al-Aqeel}
\affiliation{Department of Physics, University of Liverpool, Oliver Lodge Building, Oxford Street, L697ZE, Liverpool, UK}
\affiliation{Department of Physics, College of Science, Al-Imam Mohammad Ibn Saud Islamic University (IMISU), Riyadh, 11623, Saudi Arabia}
\author{T.~Ando}
\affiliation{Department of Physics, The University of Tokyo, 7-3-1 Hongo, Bunkyo-ku, Tokyo 113-0033, Japan}
\author{H.~Baba}
\affiliation{RIKEN Nishina Center, 2-1 Hirosawa, Wako, Saitama 351-0198, Japan}
\author{S.~Bae}
\author{S.~Choi}
\affiliation{Department of Physics and Astronomy, Seoul National University, Seoul, 08826, Republic of Korea}
\affiliation{Institute for Nuclear and Particle Astrophysics, Seoul National University, Seoul, 08826, Republic of Korea}
\author{P.~Doornenbal}
\affiliation{RIKEN Nishina Center, 2-1 Hirosawa, Wako, Saitama 351-0198, Japan}
\author{J.~Ha}
\affiliation{Department of Physics and Astronomy, Seoul National University, Seoul, 08826, Republic of Korea}
\author{L.~Harkness-Brennan}
\affiliation{Department of Physics, University of Liverpool, Oliver Lodge Building, Oxford Street, L697ZE, Liverpool, UK}
\author{T.~Isobe}
\affiliation{RIKEN Nishina Center, 2-1 Hirosawa, Wako, Saitama 351-0198, Japan}
\author{P.~R.~John}
\altaffiliation[Present address: ]{Institut f\"ur Kernphysik Technische Universit\"at Darmstadt Germany}
\affiliation{Dipartimento di Fisica e Astronomia dell' Universit\`a di Padova, Padova I-35131, Italy}  
\affiliation{INFN, Sezione di Padova, Padova I-35131, Italy} 
\author{D.~Kahl}
\affiliation{School of Physics and Astronomy, The University of Edinburgh, James Clerk Maxwell Building,  Edinburgh EH9 3FD, United Kingdom}
\author{G.~Kiss}
\altaffiliation[Present address: ]{Institute for Nuclear Research, H-4026, Debrecen, Bem ter 18/c, Hungary}
\affiliation{RIKEN Nishina Center, 2-1 Hirosawa, Wako, Saitama 351-0198, Japan}
\author{I.~Kojouharov}
\affiliation{GSI Helmholtzzentum f\"ur Schwerionenforschung GmbH, Darmstadt, Germany}
\author{N.~Kurz}
\affiliation{GSI Helmholtzzentum f\"ur Schwerionenforschung GmbH, Darmstadt, Germany}
\author{M.~Labiche}
\affiliation{STFC Daresbury Laboratory, UK}
\author{K.~Matsui}
\affiliation{Department of Physics, The University of Tokyo, 7-3-1 Hongo, Bunkyo-ku, Tokyo 113-0033, Japan}
\author{S.~Momiyama}
\affiliation{Department of Physics, The University of Tokyo, 7-3-1 Hongo, Bunkyo-ku, Tokyo 113-0033, Japan}
\author{D.~R.~Napoli}
\affiliation{INFN, Laboratori Nazionali di Legnaro, Legnaro (Padova), Italy}
\author{M.~Niikura}
\affiliation{Department of Physics, The University of Tokyo, 7-3-1 Hongo, Bunkyo-ku, Tokyo 113-0033, Japan}
\author{C.~Nita}
\affiliation{ Horia Hulubei National Institute for Physics and Nuclear Engineering (IFIN-HH), Bucharest, Romania}
\author{Y.~Saito}
\affiliation{RIKEN Nishina Center, 2-1 Hirosawa, Wako, Saitama 351-0198, Japan}
\author{H.~Sakurai}
\affiliation{Department of Physics, The University of Tokyo, 7-3-1 Hongo, Bunkyo-ku, Tokyo 113-0033, Japan}
\affiliation{RIKEN Nishina Center, 2-1 Hirosawa, Wako, Saitama 351-0198, Japan}
\author{H.~Schaffner}
\affiliation{GSI Helmholtzzentum f\"ur Schwerionenforschung GmbH, Darmstadt, Germany}
\author{P.~Schrock}
\affiliation{Center for Nuclear Study, University of Tokyo, Hongo, Bunkyo-ku, Tokyo 113-0033, Japan}
\author{C.~Stahl}
\affiliation{Institut f\"ur Kernphysik, Technische Universit\"at Darmstadt, Germany}
\author{T.~Sumikama}
\affiliation{RIKEN Nishina Center, 2-1 Hirosawa, Wako, Saitama 351-0198, Japan}
\author{V.~Werner}
\affiliation{Institut f\"ur Kernphysik, Technische Universit\"at Darmstadt, Germany}
\author{W.~Witt}
\affiliation{Institut f\"ur Kernphysik, Technische Universit\"at Darmstadt, Germany}
\affiliation{GSI Helmholtzzentum f\"ur Schwerionenforschung GmbH, Darmstadt, Germany}
\author{P.~J.~Woods}
\affiliation{School of Physics and Astronomy, The University of Edinburgh, James Clerk Maxwell Building,  Edinburgh EH9 3FD, United Kingdom}

\begin{abstract}
  Neutron-rich nuclei in the vicinity of the $N=40$ island of inversion are characterized by shell evolution and exhibit deformed ground states. In several nuclei isomeric states have been observed and attributed to excitations to the intruder neutron $1g_{9/2}$ orbital. In the present study we searched for isomeric states in nuclei around $N=40$, $Z=22$ produced by projectile fragmentation at RIBF. Delayed $\gamma$ rays were detected by the EURICA germanium detector array. High statistics data allowed for an updated decay scheme of \nuc{60}{V}. The lifetime of an isomeric state in \nuc{64}{V} was measured for the first time in the present experiment. A previously unobserved isomeric state was discovered in \nuc{58}{Sc}. The measured lifetime suggests a parity changing transition, originating from an odd number of neutrons in the $1g_{9/2}$ orbital. The nature of the isomeric state in \nuc{58}{Sc} is thus different from isomers in the less exotic V and Sc nuclei.
\end{abstract}

\date{\today}
\maketitle

\section{Introduction}\label{intro}
In the region of neutron-rich nuclei around $N=40$ the occurrence of isomeric states is a consequence of the structure evolution in this region. The valence space consists of the negative parity proton and neutron $fp$ orbitals and, in addition, for the neutrons the positive parity $\nu 1g_{9/2}$ orbital. Above $N=50$, the $s$ and $d$ orbitals also contribute.  
In \nuc{67}{Ni}, at $N=39$, a $9/2^+$ isomer at 1007~keV~\cite{pawlat95} originates from the promotion of a particle above the $N=40$ harmonic oscillator shell closure occupying the  $\nu 1g_{9/2}$ orbital~\cite{diriken14}. Systematic investigations of isomeric states at and beyond $N=40$~\cite{grzywacz98} found many more cases arising from one or two-particle configurations in the $\nu 1g_{9/2}$ orbital.
Below Ni, the neutron-rich Fe and Cr isotopes show enhanced collectivity and deformation~\cite{crawford13}. In addition to spin-gap isomers, also shape isomers are thus expected. Surveys over broad regions of nuclei found several isomers~\cite{daugas10,kameda12}, but the interpretation remains a challenge since theoretical shell model calculations need to involve a large number of shells in order to describe the structure of nuclei around \nuc{68}{Ni}. Particularly, many isomers are found in odd-odd systems, where the coupling of protons and neutrons leads to multiplets of states. This can result in states with a large difference in angular momentum and/or small difference in excitation energy, hindering the decay rate.
Experimentally, in many cases not even the spins of the ground states are known, making the interpretation challenging.

Here, we present a new study on isomeric states in the vicinity of $N=40$ and $Z=22$. We clarify the decay scheme of \nuc{60\text{m}}{V}, which was previously suggested as a cascade of two states~\cite{daugas10}, provide a first lifetime measurement of a known isomeric state in \nuc{64}{V}, and report on the discovery of a new isomeric state in \nuc{58}{Sc}. Data from the same experiment for the Ti isotopes has already been published elsewhere~\cite{wimmer19}. The proposed level schemes are compared to results from large-scale shell model calculations using the LNPS interaction~\cite{lenzi10}. 

\section{Experiment}\label{exp}
The experiment was performed at the Radioactive Isotope Beam Factory (RIBF), operated by RIKEN Nishina Center and CNS, University of Tokyo. Neutron-rich exotic nuclei were produced by fragmentation of a \nuc{238}{U} beam incident on a Be primary target (thickness 4~mm) at a beam energy of 345~$A$MeV. This way, nuclei in the vicinity and beyond $N=40$ could be produced efficiently. These were separated and analyzed in the BigRIPS fragment separator~\cite{kubo12} and transported to the experimental station at the F11 focus of the ZeroDegree spectrometer. The particle identification was achieved by measurements of the magnetic rigidity $B\rho$, the time-of-flight, and the energy loss. Two separate settings focused on \nuc{64}{V} and \nuc{60}{Ti} and the corresponding particle identification plots are shown in Fig. 1 of Ref.~\cite{wimmer19}. The nuclei were implanted in the Advanced Implantation Detector Array (AIDA)~\cite{griffin15} surrounded by the the high-purity Ge EUroball RIKEN Cluster Array (EURICA)~\cite{soederstroem13}. EURICA consists of 84 high-purity Ge crystals arranged in 12 clusters of 7 detectors sharing one cryostat. The full energy peak efficiency of the array amounted to about 10\% at 1332~keV. Three data acquisition systems recorded independently the data of BigRIPS, AIDA, and EURICA. For the analysis presented here, the data were correlated offline by means of a common synchronized time-stamp.

\section{Results}\label{sec:results}
The results for the even-odd Ti nuclei have already been presented in Ref.~\cite{wimmer19}. Here, we present a revised level scheme for \nuc{60}{V}, extract the lifetime of a previously observed isomer in \nuc{64}{V}, and report on the discovery of a new isomer in \nuc{58}{Sc}.

\subsection{Internal decay of \nuc{60}{V}} \label{sec:v60}
\begin{figure}[h]
\includegraphics[width=\columnwidth]{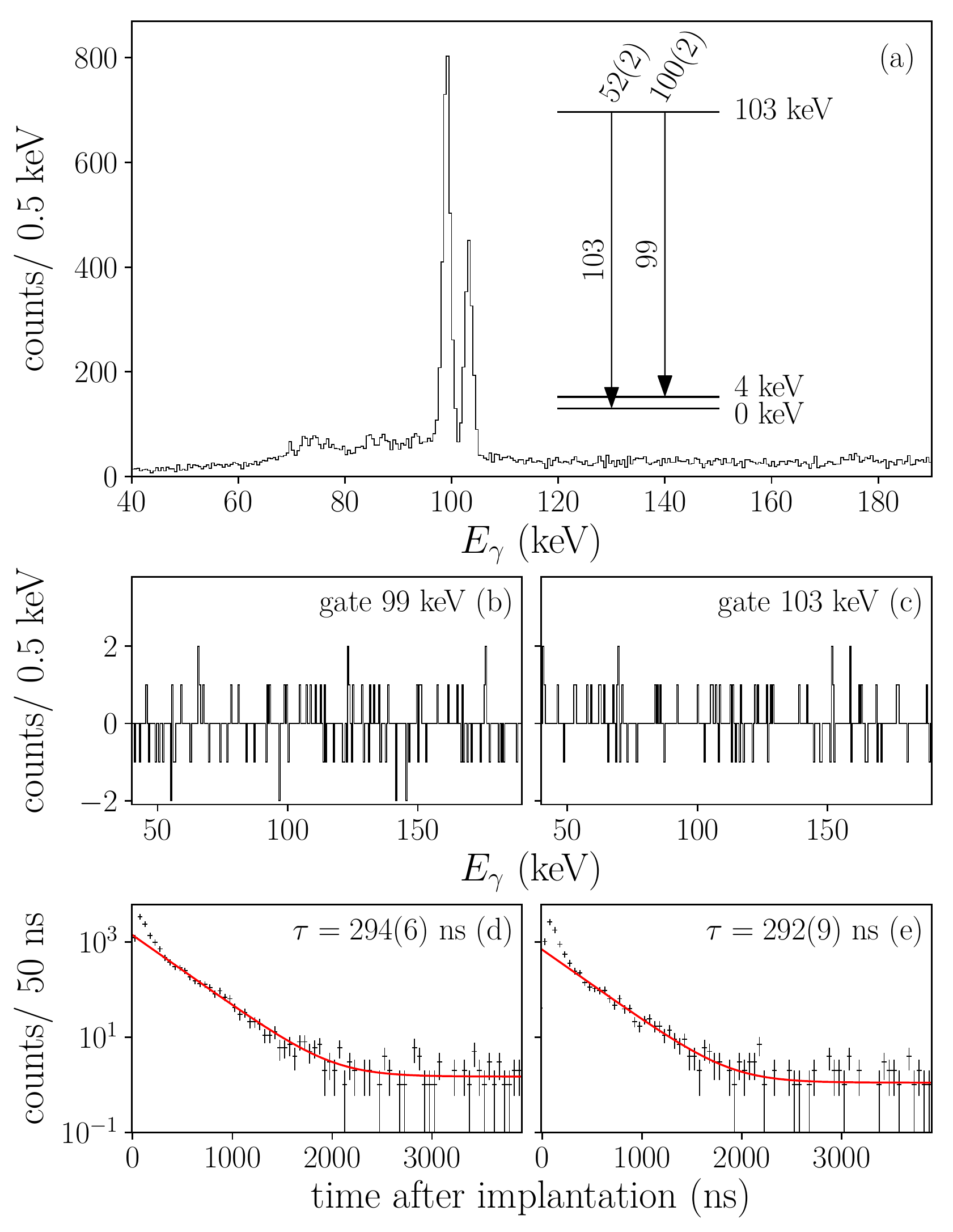}
\caption{Decay of the isomeric state in \nuc{60}{V}. Panel (a) shows the $\gamma$-ray energy spectrum in delayed coincidence with the implantation of \nuc{60}{V}. Panel (b) and (c) show the coincidence spectrum gated on each of the transitions, and panels (d) and (e) show the decay time curves together with a fit. The spectra in panels (a-c) exclude the prompt $\gamma$ flash which dominates the spectra for $t<300$~ns. The fits in panels (d) and (e) also exclude early times.}
\label{fig:v60}
\end{figure}

The isomeric decay of \nuc{60}{V} was first measured at GANIL, two transitions at 103.2 and 98.9 keV were observed~\cite{daugas99,daugas10}. The measured half-lives of 13(3) and 320(90)~ns (corresponding to lifetimes $\tau = 19(4)$ and $462(130)$~ns, respectively) suggested a sequential decay from a state at 202.1 keV through a state at 103.2~keV. The longer-lived isomer was suggested to feed the 13(3)~ns isomer, whose lifetime was extracted from a subtraction of the two decay time distributions~\cite{daugas99}. The proposed multipolarities of $E2$ for the 99~keV transition and mixed $M1$ and $E2$ transition for the 103~keV transition suggested spins and parities of $(2^+)$ and $(4^+)$ for the two states. In a previous experiment performed at the RIBF, a half-life of $229^{+25}_{-23}$~ns ($\tau = 330^{+36}_{-33}$~ns) was extracted from the sum of the decay time distributions of the 99.7 and 104~keV transitions~\cite{kameda12}. The $\gamma$-ray energy spectrum measured in the present work in delayed coincidence with identified \nuc{60}{V} ions is shown in Fig.~\ref{fig:v60} (a).
In order to exclude the prompt $\gamma$ flash originating from atomic processes, like bremsstrahlung of the ions slowing down in the silicon detectors of AIDA, as well as from secondary particles, a time gate excluding events within 300~ns of the implantation was applied.
Two transitions at 99.1(1) and 103.2(1)~keV were observed in delayed coincidence with \nuc{60}{V} identified in BigRIPS. Statistics was sufficient to construct $\gamma$-$\gamma$ coincidence matrices. The gated spectra, shown in Fig.~\ref{fig:v60} (b) and (c) for the 99 and 103~keV transitions, respectively, indicate that the two transitions proceed in parallel, and not in coincidence as previously suggested~\cite{daugas99,daugas10}. Based on the number of counts in the singles spectrum shown in Fig.~\ref{fig:v60} (a) and the efficiency of the EURICA array, we would expect several hundred counts in the coincidence spectra, when including all events from 300 ns to 5 ms after the implantation of \nuc{60}{V}.
Furthermore, the decay curves were analyzed, and the resulting lifetimes $\tau = 294(6)$ and 292(9)~ns for the two transitions are identical within their uncertainties. The intensity ratio of the two transitions also suggests a parallel decay as the relative intensity of the 99~keV transition is higher than the one of the 103~keV line which is at variance with the cascade decay proposed in Ref.~\cite{daugas99,daugas10}. The suggested level scheme is shown in Fig.~\ref{fig:v60} (a), a new state at an excitation energy of 4~keV is suggested from the present data. Such a state is likely to be isomeric and will decay by $\beta$ decay. Indeed, a $\beta$ decaying isomer was suggested, based on the two different $\beta$ decay lifetimes obtained from different experiments~\cite{sorlin03}. Weisskopf estimates indicate that both transitions are of $E2$ character with $B(E2) = 140(5)$ and $60(3)$~e$^2$fm$^4$, respectively, for the decays to the first excited, and to the ground state.

\subsection{Isomeric decay of \nuc{64}{V}} \label{sec:v64}
\begin{figure}[h]
\includegraphics[width=0.5\textwidth]{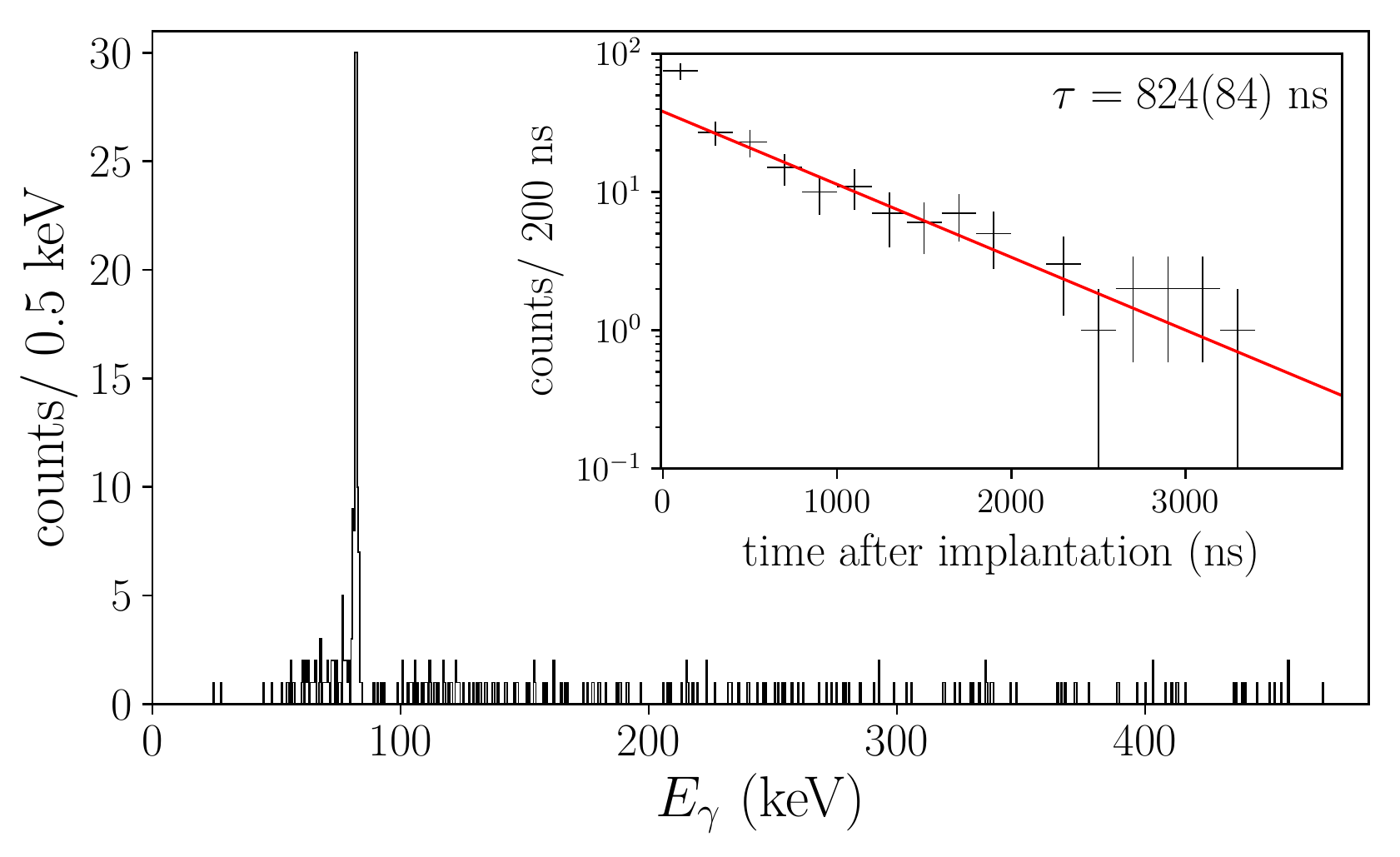}
\caption{Decay of the isomeric state in \nuc{64}{V}. The inset shows the decay time distribution with a fit using an exponential function.}
\label{fig:v64}
\end{figure}

A $\gamma$-ray originating from the isomeric decay of \nuc{64}{V} had been previously observed at NSCL. The transition energy was determined to be 81.0(7)~keV, but for the half-life only an upper limit of 1~$\mu$s could be obtained~\cite{suchyta14}. The present measurement of $\gamma$ rays in delayed coincidence with \nuc{64}{V} is shown in Fig.~\ref{fig:v64}. 
The energy of the $\gamma$ ray was measured to be 82.0(3)~keV and the lifetime 824(84)~ns is in agreement with the previously determined upper limit. The decay lifetime is consistent with a $E2$ transition with a reduced transition probability of $147(17)$~e$^2$fm$^4$.

\subsection{New isomer in \nuc{58}{Sc}} \label{sec:sc58}
\begin{figure}[h]
\includegraphics[width=0.5\textwidth]{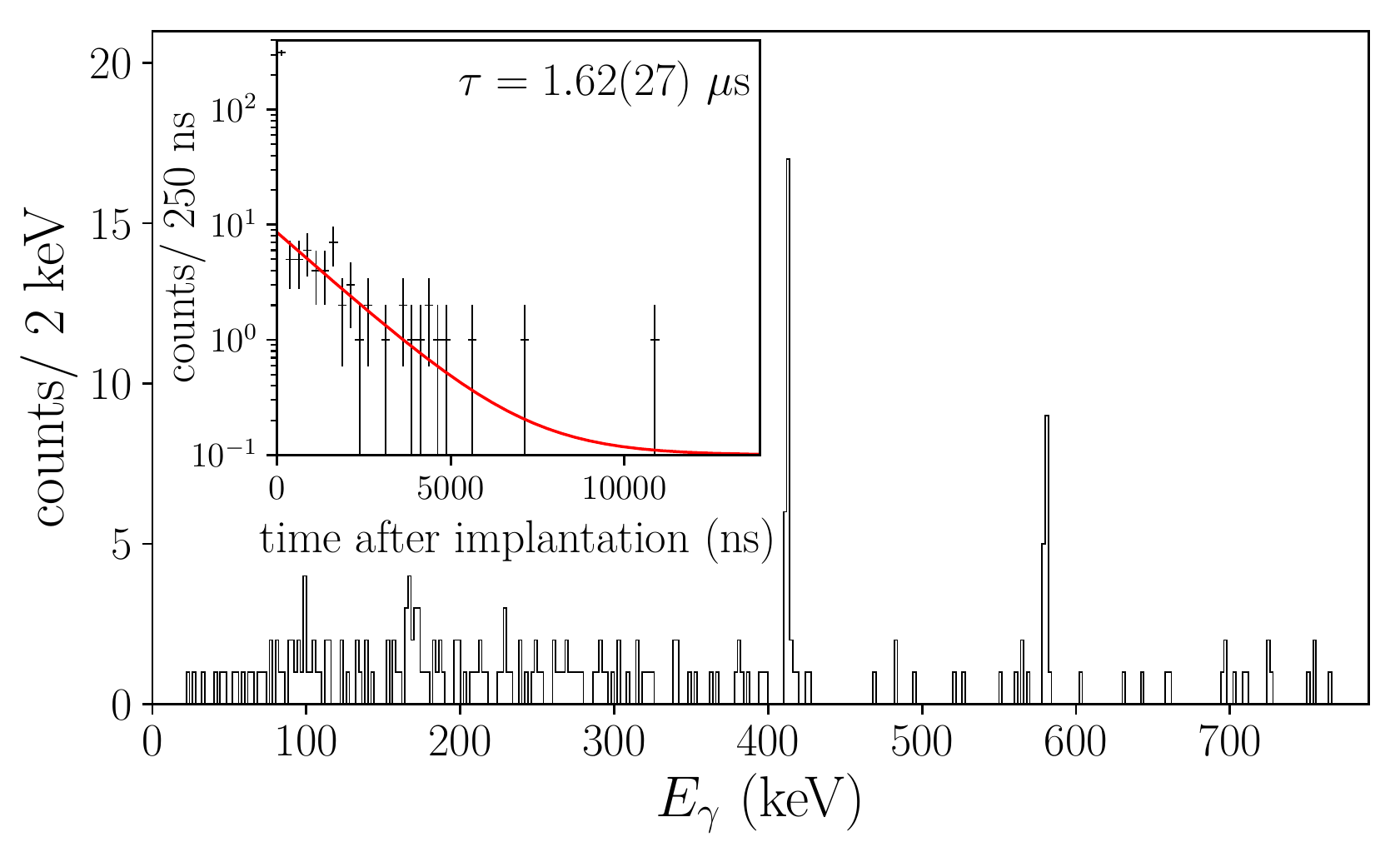}
\caption{Decay of the isomeric state in \nuc{58}{Sc}. Two transitions are clearly observed in the spectrum and indication for another one at 167~keV matches their difference. The inset shows the decay time distribution gated on both the 413 and 580~keV transitions.}
\label{fig:sc58}
\end{figure}

A new isomeric state in \nuc{58}{Sc} was discovered in the present experiment. The $\gamma$-ray energy spectrum is shown in Fig.~\ref{fig:sc58}.
Two transitions at 412.5(3) and 580.0(5)~keV were observed for the first time. The decay curves gated on these transitions yield lifetimes of 1.71(35) and 1.12(30)~$\mu$s for the 413 and 580~keV transitions, respectively. These are consistent within their uncertainties, suggesting that the two transitions originate from the same level, or are emitted in a cascade. The sum of the two time distributions which is shown in Fig.~\ref{fig:sc58} results in a lifetime of 1.62(27)~$\mu$s. The expected coincidence rate is about two events if the 413 and 580~keV transitions are in coincidence, one such event has been observed. The intensity ratio of the two transitions, however, is not in agreement with the relative efficiency, suggesting a parallel decay. 
In addition, an excess of counts at low energies suggest a peak at 167(1)~keV. This energy matches the differences between the 413 and 580 keV transitions, and thus a branch from a level at 580~keV to a state at 167~keV could be suggested. 
However, due to limited statistics, the level scheme of \nuc{58}{Sc} cannot be constructed without ambiguities. The two scenarios will be discussed in Section~\ref{sec:discu}.

\section{Discussion}\label{sec:discu}
\begin{figure}[t!]
  \includegraphics[width=\columnwidth]{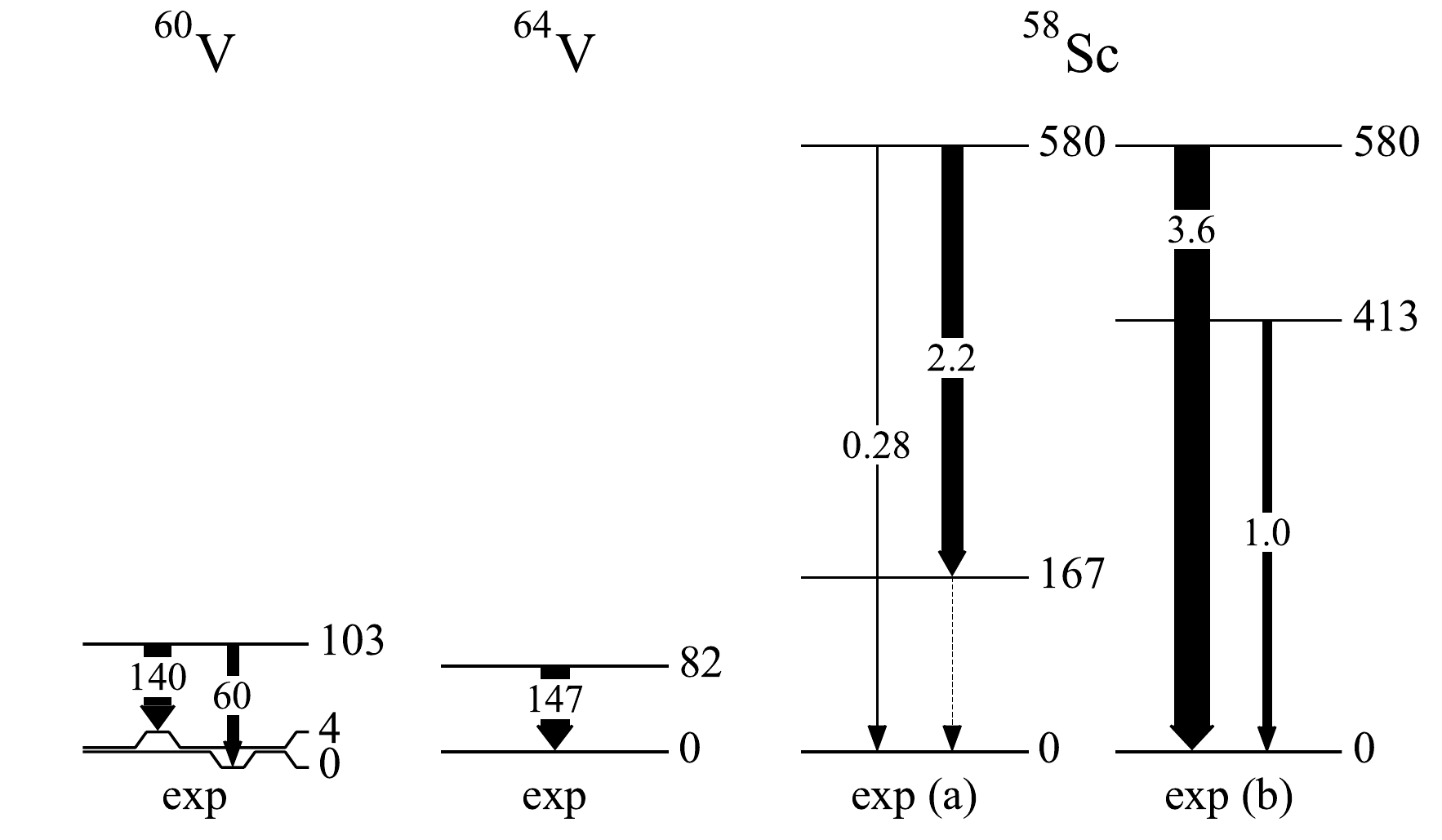}%
\caption{Level schemes of the three nuclei discussed in the present work. States are labeled with their excitation energies in keV. The arrows indicate the observed transitions with their respective reduced transition probabilities in units of e$^2$fm$^4$ for the $E2$ decays in \nuc{60,64}{V} and $\mu^2$fm$^2$ for the case of \nuc{58}{Sc} with $M2$ transitions inferred from the lifetime measurements.}
\label{fig:level}
\end{figure}

Table~\ref{tab:results} summarizes the observed isomeric transitions and the extracted lifetimes $\tau$.
\begin{table*}
\caption{Isomers observed in the present work. The transition energies, mean lifetimes, and $\gamma$-ray intensities were obtained in this work. The transition multipolarities are inferred based on comparison to Weisskopf estimates. The energies and lifetimes are also compared to previous experiments. In the case of \nuc{58}{Sc}, the level scheme could not be constructed, two scenarios are proposed, (a) an isomeric state at 580~keV decaying by two branches, and (b) two isomeric states. For details, see text.}
\label{tab:results}       
\begin{tabular}{l|rrrrrrr|rrr}
\hline
  nucleus &\multicolumn{7}{c|}{this work} & \multicolumn{2}{c}{previous work}\\
          & $E_\gamma$ (keV) & $\tau$ (ns) & $I_\gamma$  &$\pi\lambda$&$\alpha$~\cite{kibedi08}&\multicolumn{2}{c|}{$B(\pi\lambda)$ }& $E_\gamma$ (keV) & $\tau$ (ns) & Ref.\\
  \hline
  \nuc{60}{V}  &  99.1(1) & 294(6)    & 100(2)  & $E2$ & 0.392(6)  & 140(5)~e$^2$fm$^4$ & $10.0(4)$ W.u. &  98.9 &   19(4)  & \cite{daugas10} \\
               &          &           &         &      &           &                    &                &  99.7 &   -      & \cite{kameda12} \\
               & 103.2(1) & 292(9)    &  52(2)  & $E2$ & 0.335(5)  &  60(3)~e$^2$fm$^4$ &  $4.3(2)$ W.u. & 103.2 & 462(130) & \cite{daugas10} \\
               &          &           &         &      &           &                    &                & 104.0 & $330^{+36}_{-33}$     & \cite{kameda12} \\
  \hline
  \nuc{64}{V}  &  82.0(3) & 824(84)   &         & $E2$ & 0.820(17) & 147(17)~e$^2$fm$^4$& $9.7(11)$ W.u. & 81.0(7) & -     & \cite{suchyta14} \\
  \hline
  \nuc{58}{Sc} & 412.5(3) & 1620(270) & 100(24) & $M2$ & $1.80(3)\cdot10^{-3}$ & $2.2^{+0.8}_{-0.6}$~$\mu^2$fm$^2$ & $0.090^{+0.033}_{-0.024}$ W.u. & \\
   (a)         & 580.0(5) & 1620(270) & 71(26)  & $M2$ & $6.96(10)\cdot10^{-4}$  & $0.28^{+0.12}_{-0.09}$~$\mu^2$fm$^2$ & $0.011^{+0.005}_{-0.004}$ W.u. & \\
               & 167(1)   &    -      &         &      &                        &                      &\\
  \hline
  \nuc{58}{Sc} & 412.5(3) & 1710(350) & 100     & $M2$ & $1.80(3)\cdot10^{-3}$ & $3.6^{+0.9}_{-0.6}$~$\mu^2$fm$^2$ & $0.144^{+0.036}_{-0.024}$ W.u. & \\
   (b)         & 580.0(5) & 1120(300) & 100     & $M2$ & $6.96(10)\cdot10^{-4}$  & $1.0^{+0.4}_{-0.2}$~$\mu^2$fm$^2$ & $0.040^{+0.016}_{-0.008}$ W.u. & \\
  \hline
\end{tabular}
\end{table*}
For each transition, the reduced transition probabilities $B(\pi\lambda)$ have been calculated and compared to Weisskopf estimates. For the decay of the isomeric states in \nuc{60}{V} and \nuc{64}{V} these are compatible with $E2$ transitions, while $M1$ admixtures cannot be excluded.
The level schemes based on the present analysis are shown in Fig.~\ref{fig:level}.

Shell model calculations using the LNPS interaction~\cite{lenzi10} predict a high level density for \nuc{60}{V} with 10 states below an excitation energy of 500~keV. The population of the 644~keV $J^\pi=2^+$ state in \nuc{60}{Cr} in the $\beta$ decay of \nuc{60}{V} in several experiments with different ratios of the two $\beta$-decaying isomers~\cite{ameil98,sorlin99,sorlin03} suggests that they are both of low spin. The near degeneracy of the ground and first excited state is not reproduced by the calculations and no isomeric states are predicted at low excitation energy.
It should be noted that the calculated excitation energies for these odd-odd nuclei depend strongly on the two-body matrix elements of the effective interaction. Small changes, which have negligible effect for the even-even nuclei, can lead to changes in the excitation energies in the odd-odd systems by few hundred keV. Therefore, a detailed comparison between the calculated and experimental level schemes is not possible. The identification of other excited states and their decay branches to the three known states will constrain spin and parity assignments in the future.

The population of the ground and the first excited $2^+$ states in the decay of \nuc{64}{V} to \nuc{64}{Cr}~\cite{suchyta14} suggests a low spin $J\leq2$ for the ground state of \nuc{64}{V}. The shell model calculations predict a $2^+$ ground state, which is in agreement with the experiment. The predicted first excited states in \nuc{64}{V} have $J^\pi = 3^+$ and $1^+$. The calculated $E2$ and $M1$ transition probabilities, however, are not consistent with an isomeric state. At present, no other excited states are known in \nuc{64}{V} preventing us from making more detailed comparisons with the shell model calculations.

For \nuc{58}{Sc}, the data did not allow to construct the level scheme unambiguously. Two scenarios are discussed in the following. Scenario (a) assumes an isomeric state at 580~keV which decays by two branches. The 580~keV transition proceeds directly to the ground state, while the 413~keV line feeds a state at 167~keV. While the lifetimes of the two transitions are compatible within their error bars, the intensity observed for the 167~keV transition is less than for the 413~keV line after correction for efficiency. This would suggest that the 167~keV state is also long lived, with a lifetime longer than the possible correlation window of the present experiment, or with a significant $\beta$-decay branch. The study of the \nuc{58}{Sc} $\beta$ decay did not allow to discern two $\beta$-decaying states, however, the statistics were limited~\cite{gaudefroy05}. Using the lifetime of 1.62(27)~$\mu$s and the branching ratio determined from the $\gamma$-ray yield, the reduced transition probabilities suggest $M2$ transitions, albeit $E3$ transitions cannot be excluded. In this scenario, the transition from the isomer would involve thus a parity change.  
The second scenario assumes the population of two isomeric states at 580 and 413~keV. Also in this case, the measured lifetimes favor $M2$ decays, while again $E3$ cannot be fully excluded. 
Recently, isomeric states have also been reported in another experiment focused on measuring the mass of nuclei in the $N=40$ island of inversion~\cite{michimasa20}. In that study, the transitions at 413 and 580~keV have been observed, albeit with a larger relative intensity for the 580~keV $\gamma$ ray. This would favor scenario (b) with two different isomers. Statistics of that experiment were however limited and prevent firm conclusions. Two other transitions, at 181 and 247~keV were also claimed, which are not observed in the present work. The large background of the work of Ref.~\cite{michimasa20} also prevented the extraction of consistent lifetimes.

The shell model calculations predict a positive parity multiplet below $\approx 200$~keV, with negative parity states above 400~keV.
Again, the calculated high level density and the lack of experimental information on \nuc{58}{Sc} prevent us from drawing conclusions about the level scheme and favor one scenario over the other.
The calculations indicate that the low-lying states are dominated by a neutron in the $fp$ orbits, while in the excited negative-parity states have one neutron excited to the $1g_{9/2}$ orbital. This situation in \nuc{58}{Sc} is significantly different from the isotones \nuc{60}{V} and \nuc{62}{Mn}~\cite{daugas10}, where decays proceed via $E2$ (and $M1$) transitions. In \nuc{62}{Mn}, a $4^+$ $\beta$-decaying isomer at 346 keV has been proposed~\cite{heylen15,gaffney15}.
Also in the less neutron-rich Sc isotopes experimental results on isomeric and $\beta$ decays suggest positive parity for the $\gamma$ and $\beta$-decaying isomers of \nuc{54,56}{Sc} as well~\cite{crawford10}.
The parity changing transitions observed in the present work suggest a different nature of the isomeric states in \nuc{58}{Sc}. Couplings across the $N=40$ harmonic oscillator shell gap play a significant role at low excitation energy in \nuc{58}{Sc} at $N=37$. This is similar to the even-odd Ti isotopes, where the transition into the $N=40$ island of inversion starts in \nuc{59}{Ti} at $N=37$~\cite{wimmer19}. 

\section{Summary}\label{sec:sum}
In summary, new data on isomeric states in the vicinity of the $N=40$ island of inversion have been obtained. The decay scheme of the isomeric state in \nuc{60}{V} to the ground state and a very low-lying state at only 4~keV have been established using $\gamma$-$\gamma$ coincidence analysis. The decay of the isomeric state in \nuc{64}{V} has $E2$ multipolarity, based on the newly measured lifetime. Delayed $\gamma$-ray transitions in \nuc{58}{Sc} show that at least one isomeric state exists in this nucleus. However, due to limited statistics the level scheme could not be clarified. Spin and parity assignments are not possible based on the present experimental results and a comparison to shell model calculations is challenging, due to the high predicted level densities. Based on the lifetime, however, the decay is compatible with parity changing $M2$ transitions, suggesting a different nature of the \nuc{58}{Sc} isomer(s) compared to the V cases and the less exotic Sc isotopes. 

\acknowledgments
This experiment was carried out at the RIBF operated by RIKEN Nishina Center, RIKEN and CNS, University of Tokyo. We would like to thank the RIKEN accelerator and BigRIPS teams for providing the high intensity beams. 
This work has been supported by JSPS KAKENHI (Grant Nos. 25247045 and 19H00679), by the German BMBF (Grant Nos. 05P15RDFN1 and 05P19RDFN1), by the German DFG (Project No. 279384907-SFB 1245), by the STFC (UK), by the Korean National Research Foundation (grants NRF-21A20131111123, NRF-2015H1A2A1030275), by NKFIH (NN128072), and by the UNKP-20-5-DE-2 New National Excellence Program of the Ministry of Human Capacities of Hungary. K.~Wimmer acknowledges support from the Spanish Ministerio de Econom\'ia y Competitividad RYC-2017-22007. G.~Kiss acknowledges support from the Janos Bolyai research fellowship of the Hungarian Academy of Sciences. 
\bibliography{draft}

\end{document}